\documentclass[12pt, a4paper]{article}

\usepackage{multicol}

\usepackage{tikz}
\usetikzlibrary{calc,decorations.pathreplacing}
\definecolor{lightseagreen}{rgb}{0.13, 0.7, 0.67}
\definecolor{lightblue}{rgb}{0.145,0.6666,1}
\definecolor{cornflowerblue}{rgb}{0.39, 0.58, 0.93}
\definecolor{darkgreen}{rgb}{0.0, 0.44, 0.0}
\newcommand{\appr}{\overset{\cdot}{\sim}}   

\newcommand{\sbt}{\,\begin{picture}(-1,1)(-1,-2)\circle*{2}\end{picture}\ }
\usepackage{amsthm}
\usepackage{amsmath}
\usepackage{amssymb}
\usepackage{amsfonts}
\usepackage{latexsym}
\usepackage{booktabs}
\usepackage{multirow}
\usepackage{graphicx}
\usepackage{natbib}
\usepackage{array}
\usepackage{bm}

\usepackage[left = 2.5cm, right = 2.5cm, top = 3cm, bottom = 3cm]{geometry}
\usepackage[colorlinks=false, linkbordercolor=red, citebordercolor=darkgreen, urlbordercolor=darkgreen]{hyperref}

\theoremstyle{definition}

\title{Efficient computation of mixture confidence sequences  in generalized linear models}

\author{Claudia Di Caterina \\ \texttt{claudia.dicaterina@univr.it} \smallskip \\
	Department of Economics, University of Verona \\ 
	37129 Verona, Italy 
	\bigskip \\
	Luigi Pace\\ \texttt{luigi.pace@uniud.it}\smallskip\\
	Department of Economics and Statistics, University of Udine\\
	33100 Udine, Italy
	\bigskip\\
	Alessandra Salvan \\ \texttt{alessandra.salvan@unipd.it} \smallskip \\
	Department of Statistical Sciences, University of Padova \\
	35121 Padova, Italy  
	\bigskip \\
	Nicola Sartori \\ \texttt{sartori@stat.unipd.it} \smallskip \\
	Department of Statistical Sciences, University of Padova \\
	35121 Padova, Italy  	
}

\begin{document}
	
	\maketitle
	
	\begin{abstract}\noindent
	
			\noindent
Classical confidence intervals, when repeatedly obtained on accumulating data at different sample sizes, produce contradictory inferences with high probability. We propose a simple and efficient strategy for computing, instead,  mixture confidence sequences for regression coefficients in generalized linear models under this sequential framework. Simulations demonstrate the computational convenience of our approach  and the importance of drawing inferential conclusions based on these anytime-valid tools when  observations become available in batches over time. The usefulness of the proposed procedure is also shown in the analysis of streaming data from the American National Automotive Sampling System.

	\end{abstract}
	
	timation; Streaming data.

	\section{Introduction}
	When inference is performed on accumulating data, repeatedly  applying  fixed-sample-size methods can seriously undermine replicability. In the era of big data, streaming data sets are the norm rather than the exception. Such data may arise, for instance,  with  A/B testing in on-line randomized experiments \citep{johari22, Larsen24} or safety surveillance \citep{nelson15, luoEtAl24}.  In these contexts, it is especially important to deliver safe inferences which remain valid at any point in time during the analysis.

If the procedure of interest is a confidence interval, standard methods  can lead to contradictory conclusions. It is almost sure that non-overlapping confidence intervals  arise along the way, see e.g. \citet{pace20}. 
On the other hand, confidence sequences \citep{darling67, robbins70, lai76, howard21, ramdas23} meet the coverage requirements even  under arbitrary enlargements of the sample. They provide a uniform coverage guarantee and  allow for continuous monitoring and optional stopping. As a result,   confidence sequences 
constructed at different sample sizes are compatible, meaning that they overlap with high  probability \citep{pace20}.  

Approximate mixture confidence sequences for a scalar parameter of interest can be based on an asymptotically normal estimator \citep{pace20, pace23}.  \cite{waudby-smith:24} propose asymptotic confidence sequences based on time-uniform central limit theory.
Here we focus on mixture confidence sequences for regression coefficients in generalized linear models.  In order to ensure computational  efficiency,  we propose to use the renewable estimation algorithm of \cite{luosong20}, which is convenient in terms of computing time, memory storage and/or previous data availability (cf. Section \ref{dicaterina:secCS} for more details).
Mixture confidence sequences are then built based on the sequential update and on the asymptotic normality of the renewable estimator. We also investigate how the empirical tuning of the mixing distribution affects the accuracy of the resulting confidence sequences.	
	
The remainder of this paper is organized as follows. Section 	\ref{dicaterina:secCS} sets the notation in the sequential framework, presents the renewable estimation procedure, and reviews the relevant literature on mixture confidence sequences. 
In Section \ref{sec:simu}, we combine the two approaches and use simulations to demonstrate how {mixture confidence sequences for generalized linear models are easy to obtain when many data batches accrue over time, and deliver more reliable inferences than usual Wald intervals.}
Section \ref{sec:NASS} contains an application of the proposed strategy to the National Automotive Sampling System data, and Section \ref{sec:final} concludes with some final remarks.

	%
	%
	
	\section{Renewable estimation and mixture confidence sequences for streaming data sets}
	\label{dicaterina:secCS}
	
	Assume that  data $y_i$, $i=1,\ldots,n$, are realizations of independent random variables $Y_1,\ldots,Y_n$ following a generalized linear model (GLM) with density $p(y_i; x_i, \beta, \phi)$, where $x_i$ is a $p$-dimensional vector of covariates,  $\beta= (\beta_1,\dots, \beta_p)^\top$ is a vector of  coefficients and $\phi > 0$ is a dispersion parameter. The $n\times p$ model matrix $X$ with $i$th row $x_i^\top$ is assumed to have full rank $p<n$. Throughout, $\beta_1$ is an intercept term, so that the first  column of $X$ is the unit vector. We let $\mu_i=E(Y_i|x_i)= g(\beta^\top x_i)$ for some known link function $g(\cdot)$. 
	
	We consider the same setting and notation as in \cite{luosong20} where data arrive in batches $D_b=\{y_b^{n_b}, X_b\}$,  $b=1,\dots,B$,  with $y_b^{n_b}$ the $b$th response vector with $n_b$ observations, and $X_b$ the corresponding $n_b\times p$ submatrix of $X$. In general, it is not possible to  update the maximum likelihood estimate in a GLM 
	 sequentially, based solely on 
	 $D_b$ and  the previous 
	 estimate. 
	The incremental updating algorithm by \cite{luosong20}, called renewable estimation, allows the maximum likelihood estimator to be updated as new batches become available, using current data and summary statistics of historical data. 
	
	%
	Let $y^{(N_b)}$ denote the data $(y_1^{n_1},\ldots, y_b^{n_b})$ up to the $b$th batch, realization of   $Y^{(N_b)}=(Y_1^{n_1},\ldots, Y_b^{n_b})$, $b=1,\ldots,B$, $N_b=\sum_{j=1}^b n_j$.   Thus, the total sample size is $n=N_B$. 
	The goal is to  estimate
	$\mu_i=E(Y_i|x_i)= g(x_i^\top\beta)$,  $i = 1,\dots,N_b$, by  fitting  sequentially a GLM on the first $b$ aggregated batches, $D^{(N_b)}=\{D_1,\dots,D_b\}$, every time a new batch $D_b$ is observed.
	Denote by $U_b(D_b;\beta)=\sum_{i \in D_b}  \nabla_\beta  \,\log p(y_i; x_i, \beta, \phi)$ 
	the score function
	and by $J_b(D_b;\beta)=-\nabla_\beta U_b(D_b;\beta)$ the observed information of the $b$th batch. 
	The renewable estimator $\tilde\beta_b$ proposed by \citet{luosong20} is the solution to the incremental estimating equation 
	\begin{equation}\label{luo_song_10}
		\sum_{j=1}^{b-1}		J_j(D_j;\tilde\beta_j)(\tilde\beta_{b-1}-	\tilde\beta_{b})+
		U_b(D_b;\tilde\beta_b) = 0\,,
	\end{equation}
	where $\tilde\beta_1$ 
	equals the maximum likelihood estimator for data $D_1$. Let  the aggregated observed  information be $\tilde J_b=\sum_{j=1}^b J_j(D_j;\tilde\beta_j)$.  The iterative algorithm for solving (\ref{luo_song_10}) has 
	$r$th iteration 
	\citep[formula (11)]{luosong20} 
	\begin{equation*} \label{dicaterina:eq_incr_alg}
		\tilde\beta_b^{(r + 1)} = \tilde\beta_b^{(r)} + \{ \tilde J_{b-1} + J_b(D_b; \tilde\beta_{b-1})\}^{-1}\tilde U_b^{(r)},
	\end{equation*}
	where $\tilde U _b ^{(r)} = \tilde J_{b-1}(\tilde\beta_{b-1} - \tilde\beta_b^{(r)}) + U_b(D_b;\tilde\beta_b^{(r)})$ is the adjusted score. As noted before, the implementation of the whole procedure requires just the current data $D_b$ and summary statistics $\{ \tilde\beta_{b-1}, \tilde J_{b-1}\}$ from the historical observations.
	
	The renewable estimator $\tilde\beta_b$ is 
	consistent  and asymptotically normal in a conventional way 
	as $N_b \rightarrow \infty$ \citep[Section 4]{luosong20}, with estimated covariance matrix $\tilde V_b= \tilde\phi_b \tilde J_b^{-1}$, where  $\tilde\phi_b$ is updated iteratively from $\tilde\phi_{b-1}$ \citep[p. 77]{luosong20}. 
	Such theoretical results make the use of the Wald statistic straightforward for inference on  components of $\beta$. Yet such practice has only fixed-sample-size validity and, indeed, the empirical coverage probabilities  obtained via simulation in  \citet[Section 6]{luosong20} are referred specifically to the final sample size. 
	To take into account the streaming structure of the data generating process, it is instead natural to perform inference sequentially as batches accumulate. This requires uniform validity of inference procedures. 
	Confidence intervals with such properties are known  as confidence sequences (CSs)  \citep{darling67, robbins70, howard21} and can be repeatedly computed under continuous monitoring or at arbitrary stopping times. 
	
	Considering a generic parametric model with parameter $\theta\in\Theta\subseteq \mathbb{R}^p$, we denote by $p(y^{(N_b)};\theta )$ the density of  $Y^{(N_b)}$, where possible conditioning on covariates is understood. We assume that the support of $Y^{(N_b)}$ does not depend on $\theta$.  
	A  sequence of estimation regions based on $y^{(N_1)}, y^{(N_2)},\ldots $ and denoted by $\hat\Theta_{N_b}=\hat{\Theta}( y^{(N_b)})\subseteq\Theta$ is a $(1-\alpha)$-CS if, for every $\theta\in\Theta$, 
	$$ 
	P_\theta \left( \theta \in \cap_{b \geq 1}\hat\Theta_{N_b}   \right) \geq 1 - \alpha\,, \quad 0 < \alpha <1.
	$$ 
	For a $(1-\alpha)$-CS, the probability of observing incompatible conclusions when the sample enlarges is smaller than $\alpha$ \citep{pace20}. This means that regions in the CS overlap with high enough probability. In particular, for every $\theta\in\Theta$, 
	$$ 
	P_\theta\left(  \cap_{b \geq 1}\hat\Theta_{N_b} = \emptyset \right) \leq \alpha \,, \quad 0 < \alpha <1.
	$$ 
	Ville's inequality \citep[see, e.g.,][Section 2.5]{ramdas23} is the starting point to obtain $(1-\alpha)$-CSs. Among those, mixture CSs \citep{darling67, robbins70} have the form 
	\begin{equation}\label{dicaterina:mixture0}
		\hat\Theta_{N_b}   = \left\{ \theta \in \Theta \, : \; p(y^{(N_b)};\theta ) > \alpha   \int_\Theta p(y^{(N_b)};\theta ) \pi(\theta) \, d\theta    \right\} \, ,
	\end{equation}
	where the weight function $\pi(\theta)$ is a preset probability density over $\Theta$.
	Confidence sequences (\ref{dicaterina:mixture0})  are likelihood based, can incorporate prior information, obey the strong likelihood principle and may be seen as support sets \citep{pawel23}.
	
	Let the parameter be partitioned as  $\theta = (\psi , \lambda)$, with $\psi\in \Psi \subseteq \mathbb{R}^{p_0}$ the component of interest and $\lambda$ the  nuisance parameter. 
	From \cite{pace20}, 	a possible ($1-\alpha$)-CS for $\psi$ is obtained by projecting $\hat\Theta_{N_b} $ on $\Psi$ and can be expressed in terms of profile likelihood as 
	\begin{equation}\label{dicaterina:mixture-profile1} 
		\hat{\Psi}_{N_b} = \left\{ \psi \in \mathbb{R} \, : \;  p(y^{(N_b)};\psi, \hat{\lambda}_\psi ) > \alpha  \int_\Theta p(y^{(N_b)};\theta ) \pi(\theta) \, d\theta  \right\}\,,
	\end{equation}
	where  $\hat{\lambda}_\psi$ is the maximum likelihood estimate of $\lambda$ with  $\psi$ fixed computed using $y^{(N_b)}$. 
	
	Sequences (\ref{dicaterina:mixture-profile1}) are likely to be far more conservative than their counterpart with known $\lambda$, as discussed in \citet[Example 2]{pace20}. A viable and simple approximation is proposed in \citet{pace20} and \citet{pace23} for a scalar $\psi$. Suppose that a normal approximation  $N(\psi, s_b^2)$ is available for the distribution of the maximum likelihood estimator $\hat\psi_{b}$ calculated on data $y^{(N_b)}$, where $s_{b}^2$ is the estimated variance of $\hat\psi_b$. 
	Using a  $N(\psi_0, \tau^2_0)$ weight function, the  closed-form approximate mixture CS for $\psi$ proposed in \citet[formula (17)]{pace23} becomes
	\begin{equation}\label{dicaterina:closed-form}
		\hat\psi_b \pm  s_b \sqrt{\log \dfrac{\tau^2_0 + s_b^2}{s_b^2} + \dfrac{(\hat\psi_b- \psi_0)^2}{\tau_0^2 + s_b^2} - 2 \log \alpha}\,. 
	\end{equation}
	Approximate mixture CSs are typically very conservative. To tighten the boundaries in practice, one can 
	replace $\psi_0$ and $\tau_0^2$ in (\ref{dicaterina:closed-form}) with the estimate of $\psi$ and its squared standard error, respectively, from a preliminary batch of $n_0$ observations.   
	Although based on a fixed-sample-size normal approximation,  CSs (\ref{dicaterina:closed-form}) with pre-specified $\psi_0$ and $\tau^2_0$ have shown satisfactory uniform coverage properties in simulation studies reported in \citet{pace20} and \citet{pace23}. 
	
	\cite{waudby-smith:24} propose asymptotic CSs based on time-uniform central limit theory, guaranteeing that Gaussian approximations hold almost surely for all sample sizes simultaneously. For a sequence $Y_t$, $t=1,2,\ldots$, of independent and identically distributed observations from a distribution with mean $\mu$ and finite variance, the Gaussian mixture asymptotic $(1-\alpha)$-CS for $\mu$ \citep[][formula (8)]{waudby-smith:24} is 
	\begin{equation}\label{waudby-smith-8}
		\bar{Y}_t \pm  \dfrac{\hat\sigma_t}{\sqrt{t}} \sqrt{\dfrac{t\rho^2 + 1}{t\rho^2}\log \dfrac{t\rho^2 + 1}{\alpha^2} }\,, 
	\end{equation}
	where $\bar{Y}_t=\sum_{i=1}^t Y_i/t$ and $\hat\sigma_t^2=\sum_{i=1}^t Y_i^2/t-\bar{Y}_t^2$. \citet[][Appendix B.2]{waudby-smith:24} suggest to choose the tuning parameter $\rho$ as the minimizer of the expected length of (\ref{waudby-smith-8}) for fixed $\alpha$ and $t$. The optimal $\rho$ is the solution  to the equation
	\begin{equation}\label{opt_rho}
		\alpha^2 \exp\{t\rho\}= t\rho +1\,.
	\end{equation}
	To construct an asymptotic CS for a parameter of interest $\psi$, the result in \citet[Corollary 3.4]{waudby-smith:24} for general functional estimation can be invoked. Specifically, in (\ref{waudby-smith-8}) we can replace $\bar{Y}_t $ with  $\hat\psi_b$ and $\hat\sigma_t/\sqrt{t}$ with the  standard error $s_b$.

	
	%

	\section{Mixture confidence sequences in 
		generalized linear models} 
	\label{sec:simu}
\begin{figure}[t]
	\begin{center}
		\includegraphics[height=11.3cm]{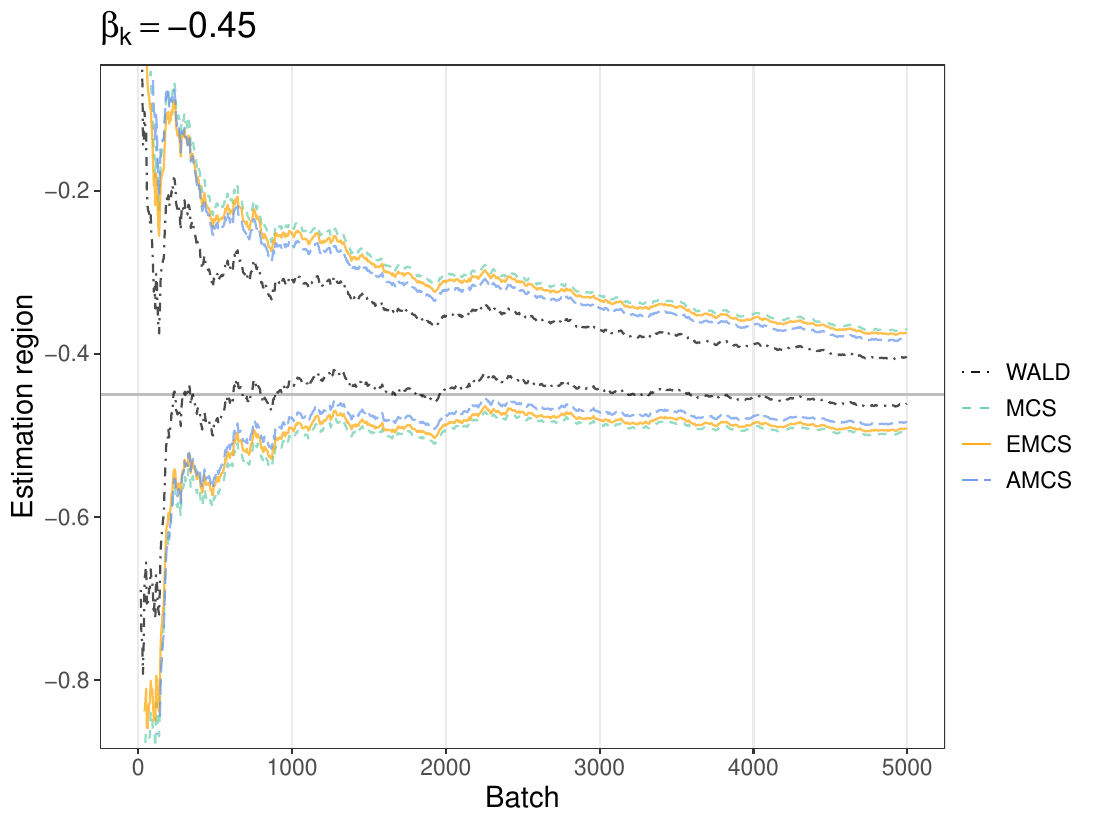}
		\caption{Estimation regions for data in $B = 5000$ batches of size $n_b=20$,  generated from a logistic regression model with $p=20$; 
			WALD (dot-dashed), MCS (dashed), EMCS (solid), AMCS (long-dashed) with $\rho$ optimized at $t=2 \times 10^4$, $\alpha=0.1$,  $n_0=200$. The gray horizontal line marks the true parameter value $\beta_{k}=-0.45$.}\label{sampleCSsBin}	
	\end{center}
\end{figure}

	Here we study the behaviour of the mixture CSs introduced in the previous section for inference about a scalar component of $\beta$ in a GLM. 
	In particular, we consider as a parameter of interest $\psi$ the  $k$th regression coefficient $\beta_k$, whose renewable estimator $\tilde\beta_{k,b}$ has approximate distribution $\tilde\beta_{k,b} \appr N(\beta_k, \tilde V_{k,b})$ for some given large sample size $N_b$. We  compute the following estimation regions: 
	\begin{itemize}
	\item[{ }] (WALD) $(1-\alpha)$-Wald confidence interval $\tilde\beta_{k,b} \pm z_{1-\alpha/2}\sqrt{\tilde{V}_{k,b}}$, where $z_q$ is the standard normal $q$ quantile;  			
	\item[{}] (MCS) approximate mixture $(1-\alpha)$-CS with $N(0,1)$ weight function, obtained computing (\ref{dicaterina:closed-form}) with  $\hat\psi_b=\tilde\beta_{k,b}$, 
	$ s_{k,b}^2=  \tilde V_{k,b}$, $\psi_0=0$, $\tau^2_0=1$;
	%
	%
	\item[{}]  (EMCS) empirical 
	approximate mixture
	 $(1-\alpha)$-CS as in (MCS) with $\psi_0$ replaced by $\hat\beta_{0,k}$ and $\tau_0^2$ replaced by $s^2_{{0},k}$,
	 the maximum likelihood estimates of $\beta_k$ and its 
	 squared standard error
	computed on a preliminary batch of $n_{0}$ observations;
	%
	%
	%
	\item[{}] (AMCS) asymptotic   mixture 
	$(1-\alpha)$-CS  as in  (\ref{waudby-smith-8}) with $\bar{Y}_t $ replaced by  $\tilde\beta_{k,b}$,  the estimated standard deviation $\hat\sigma_t/\sqrt{t}$ replaced by $\tilde V_{k,b}^{1/2}$ and using as the optimal $\rho$ the  solution to (\ref{opt_rho}) for a chosen $t$. 
	
\end{itemize}
\vspace{0.5cm}
As an illustration, in Figure \ref{sampleCSsBin} Wald confidence intervals are plotted together with CSs with $1-\alpha=0.9$ for one coefficient $\beta_{k}$ 
in a logistic regression model. 
The model has $p=20$ covariates, including the intercept. 
These were generated in R \citep{R} from a multivariate normal distribution with mean 0 and covariance matrix $\Sigma/n_0$, where $\Sigma_{kk} = 1$, $\Sigma_{kk'} =  0.5$, $k \neq k'$. Such a covariance matrix was chosen for preventing separation issues in the preliminary batch due to the magnitude of $p$ \citep{sur19}. The first ten 
covariates were made binary by setting their value to 0 if the underlying continuous value was lower than 0, and to 1 otherwise. Binary response observations with success probability $\mu_i = e^{\beta^\top x_i}/(1 + e^{\beta^\top x_i})$
were then simulated for the preliminary batch of size $n_0=200$ and the subsequent $B=5000$ batches of size $n_b=20$. For each batch $D_b$, $b=1, \dots,B$, the renewable estimator $\tilde\beta_{k,b}$ and its estimated variance $\tilde V_{k,b}$ were computed using the package \texttt{RenewGLM} \citep{renewGLM}. The optimal  $\rho$ in AMCS was computed for $t=N_B/5$,
which was found to be the most favorable value among those tried.
Wald confidence intervals fail to include the true parameter value several times along the sequence, unlike  the considered  time-uniform CSs. Among the latter, AMCS seems preferable, as it is typically shorter than both MCS and EMCS.

We  can empirically  check whether the example above reflects the actual general performance of the confidence regions in the  logistic regression framework.
Cumulative miscoverage probabilities under the null and non-null settings are estimated through simulation experiments  based on $10^4$ Monte Carlo replications. Throughout, ``null'' refers to the true parameter value used to generate the Monte Carlo samples, whereas ``non-null'' refers to any other parameter value.

\begin{figure}
	\begin{center}
		 		\includegraphics[height=13.5cm]{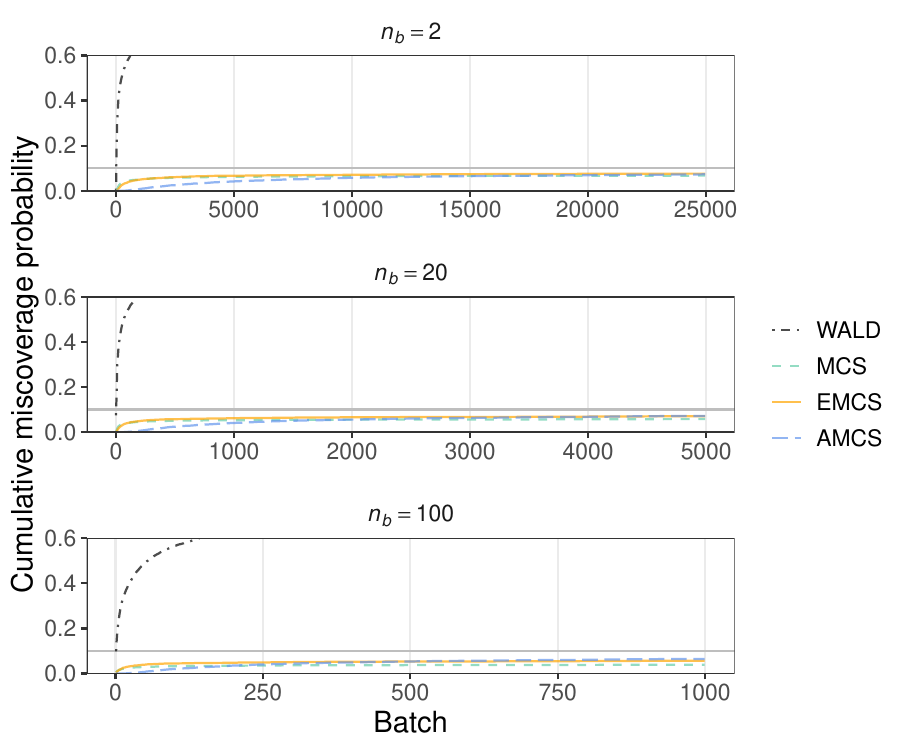}
		 		\caption{Inference about the log-odds ratio $\beta_{2}=-0.288$ in $2\times 2$ tables. Empirical cumulative null miscoverage probabilities of estimation regions with $\alpha=0.1$ for $n_b=2, B=25000$ (top), $n_b=20, B=5000$ (middle), $n_b=100, B=1000$ (bottom); WALD (dot-dashed), MCS (dashed), EMCS (solid) with $n_0=200$, AMCS (long-dashed) with $\rho$ optimized at $t=10^4$ when $n_b=2$ and $t=2 \times 10^4$ when $n_b=20, 100$. In each panel, the gray horizontal line marks the nominal value $\alpha$. }\label{sim_2x2}	
	\end{center}
\end{figure}

\begin{figure}
	\begin{center}
		 		\includegraphics[height=13.5cm]{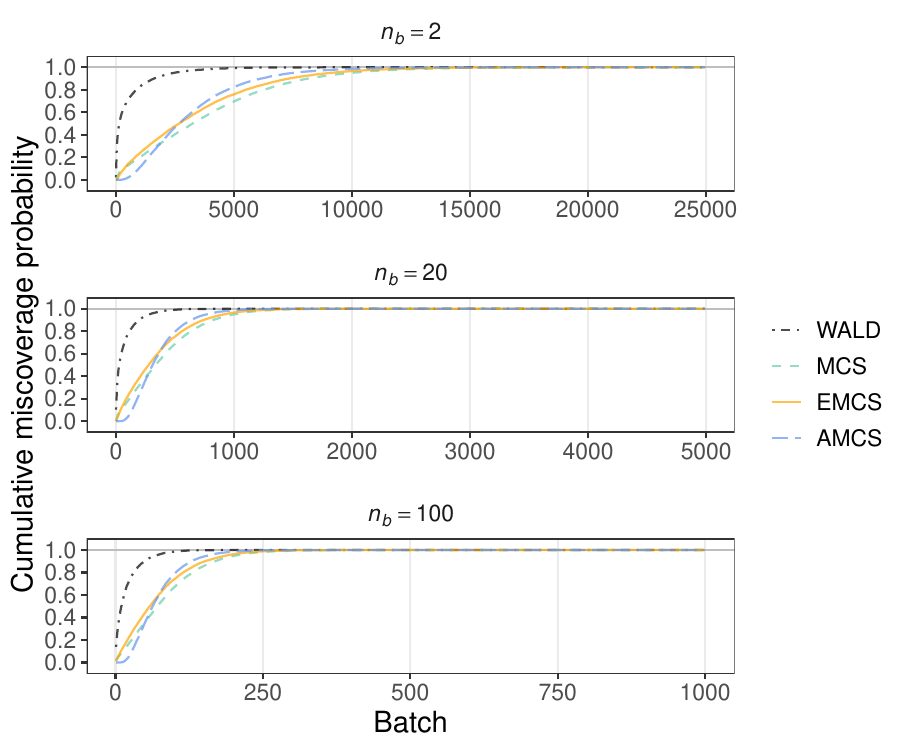}
		 		\caption{Inference about the log-odds ratio $\beta_{2}=-0.288$ in $2\times 2$ tables. Empirical cumulative non-null miscoverage probabilities, $\beta_2'=-0.45$, of estimation regions with $\alpha=0.1$ for $n_b=2, B=25000$ (top), $n_b=20, B=5000$ (middle), $n_b=100, B=1000$ (bottom); WALD (dot-dashed), MCS (dashed), EMCS (solid) with $n_0=200$, AMCS (long-dashed) with $\rho$ optimized at $t=10^4$ when $n_b=2$ and $t=2 \times 10^4$ when $n_b=20, 100$. In each panel, the gray horizontal line marks the 1 target value. }\label{sim_2x2NN}	
	\end{center}
\end{figure}

As a first experiment, we consider a single binary covariate $x_i$, giving success probability  $\mu_i = e^{\beta_1+ \beta_2 x_i}/(1 + e^{\beta_1+\beta_2 x_i})$, i.e., $\mu_i=\mu_{i1} = e^{\beta_1}/(1 + e^{\beta_1})$ when $x_i=0$ and 
$\mu_i=\mu_{i2} = e^{\beta_1+\beta_2}/(1 + e^{\beta_1+\beta_2})$ when $x_i=1$. 
We set $\mu_{i1}=0.2$ and $\mu_{i2}=0.25$, so that $\beta_2=-0.288$ is the log-odds ratio.
Batches  of $n_b \in \{2, 20, 100\}$ observations are randomly generated,  where $x_i=0$  for the first half   and $x_i=1$ for the second half. These response values are denoted by $y_{i1}$ and $y_{i2}$, respectively.  Thus, for each batch $D_b$, $y_{\sbt l, b}=\sum_{i\in D_b} y_{il}$, $l=1,2$,  
 is a realization of a binomial distribution with index 
$n_b/2$ and success probability $\mu_{il}$. For this setting, the iterative algorithm based on (\ref{luo_song_10}) is not needed because closed-form estimates, with continuity correction, are  given by
$$
\hat\beta_{2,b} =\log\frac{(y_{\sbt 1, b}+0.5)(n_b/2-y_{\sbt 2,b}+0.5)}{(n_b/2-y_{\sbt 1, b}+0.5)(y_{\sbt 2,b}+0.5)}
$$ 
and
$$
s_{2,b}^2=\frac{1}{y_{\sbt 1,b}+0.5}+\frac{1}{n_b/2-y_{\sbt 1,b}+0.5}+\frac{1}{y_{\sbt 2,b}+0.5}+\frac{1}{n_b/2-y_{\sbt 2,b}+0.5}\,,
$$
see, e.g., \citet{gart1985} and \citet[][Section 2.3]{agresti19}.
The results are summarized in Figure \ref{sim_2x2}	 for  miscoverage probabilities of the true parameter value $\beta_2=-0.288$. While the null cumulative miscoverage rate of Wald confidence intervals rapidly goes to 1, those of all CSs stay as desired  under the nominal level $\alpha$.  Figure \ref{sim_2x2NN} shows  empirical non-null cumulative miscoverage probabilities, where rates of miscoverage are computed for the 
parameter value $\beta_2'=-0.45$. In this case we see that even CSs correctly deliver cumulative probabilities that  reach 1, with 
the three curves intersecting. Given the optimal tuning parameter selected for AMCS, this appears to be the best overall estimation region. However, we found that  the choice of $\rho$ can affect
its behaviour in terms of  non-null miscoverage rate, and so the relative performance of CSs.



\begin{figure}	
	\begin{center}
		\includegraphics[height=16cm]{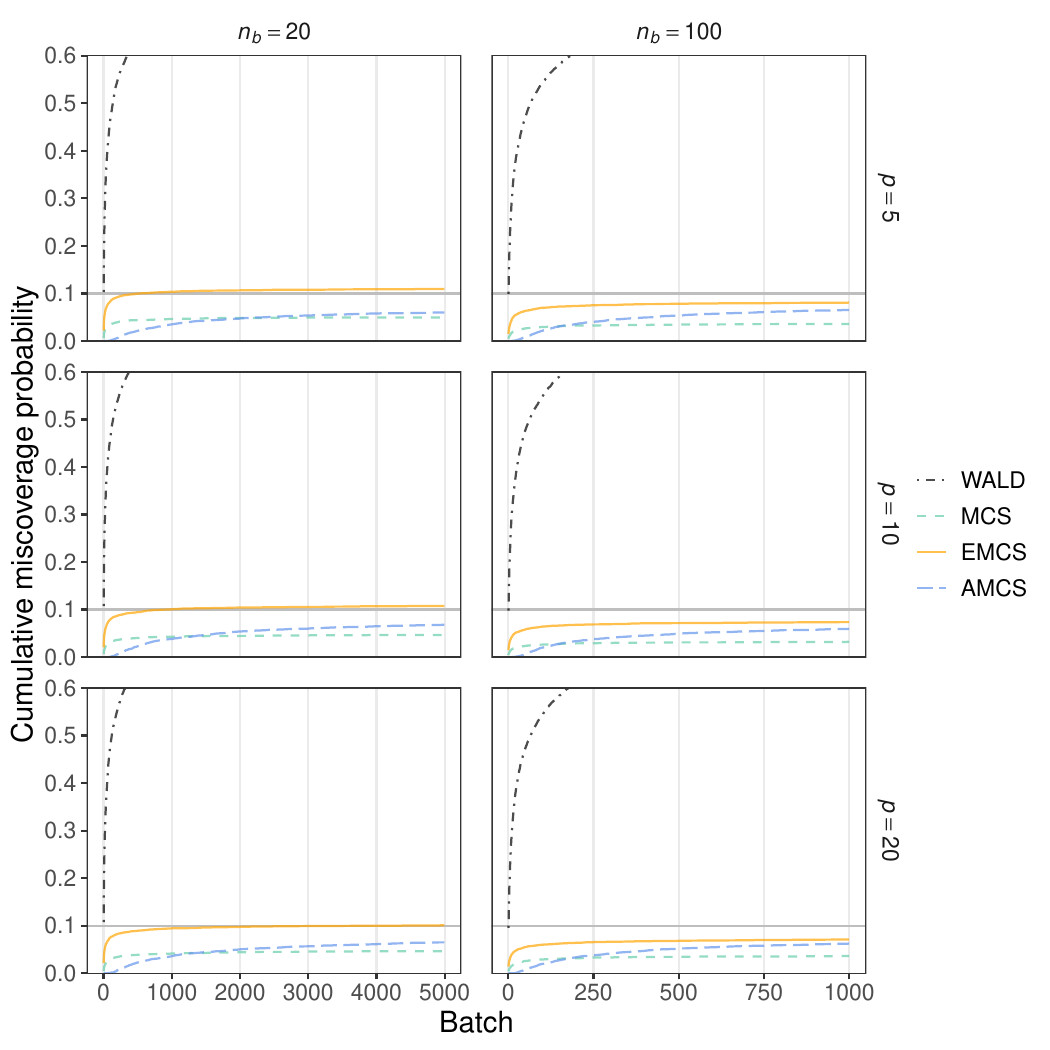}
		\caption{Inference about the coefficient $\beta_{k}=-0.45$ of a binary covariate in the logit regression model. Empirical cumulative null miscoverage probabilities of estimation regions with $\alpha=0.1$ for $p=5$ (top), $p=10$ (middle), $p=20$ (bottom), and $n_b=20$ (left), $n_b=100$ (right); WALD (dot-dashed), MCS (dashed), EMCS (solid) with $n_0=200$, AMCS (long-dashed) with $\rho$ optimized at $t=2 \times 10^4$. In each panel, the gray horizontal line marks the nominal value $\alpha$. }\label{sim_Logit_Bin}
	\end{center}
\end{figure}

\begin{figure}	
	\begin{center}
		\includegraphics[height=16cm]{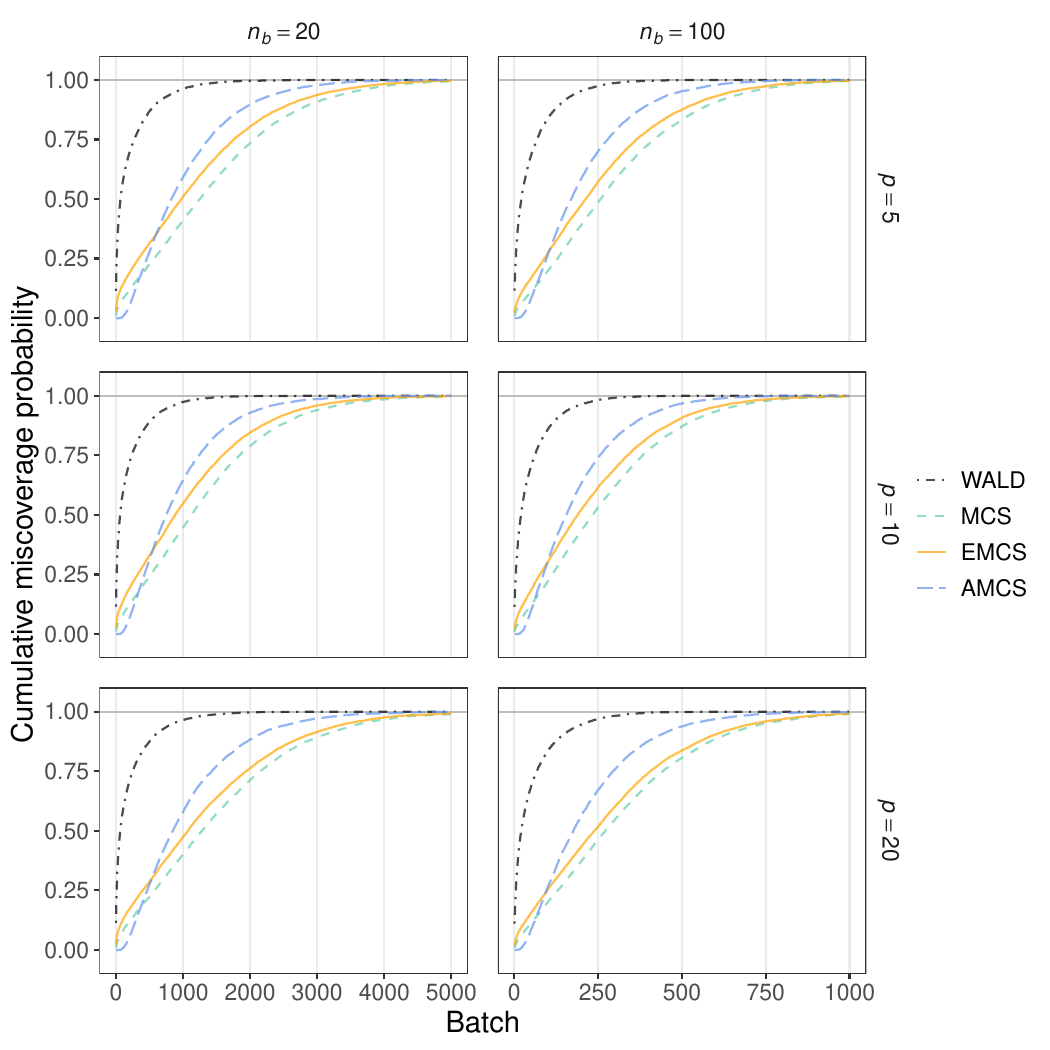}
		\caption{Inference about the coefficient $\beta_{k}=-0.45$ of a binary covariate in the logit regression model. Empirical cumulative non-null miscoverage probabilities, $\beta_k'=-0.55$, of estimation regions with $\alpha=0.1$ for $p=5$ (top), $p=10$ (middle), $p=20$ (bottom), and $n_b=20$ (left), $n_b=100$ (right); WALD (dot-dashed), MCS (dashed), EMCS (solid) with $n_0=200$, AMCS (long-dashed) with $\rho$ optimized at $t=2 \times 10^4$.  In each panel, the gray horizontal line marks the 1 target value.}\label{sim_Logit_Bin_NN}
	\end{center}
\end{figure}

\begin{figure}
	\begin{center}
		\includegraphics[height=16cm]{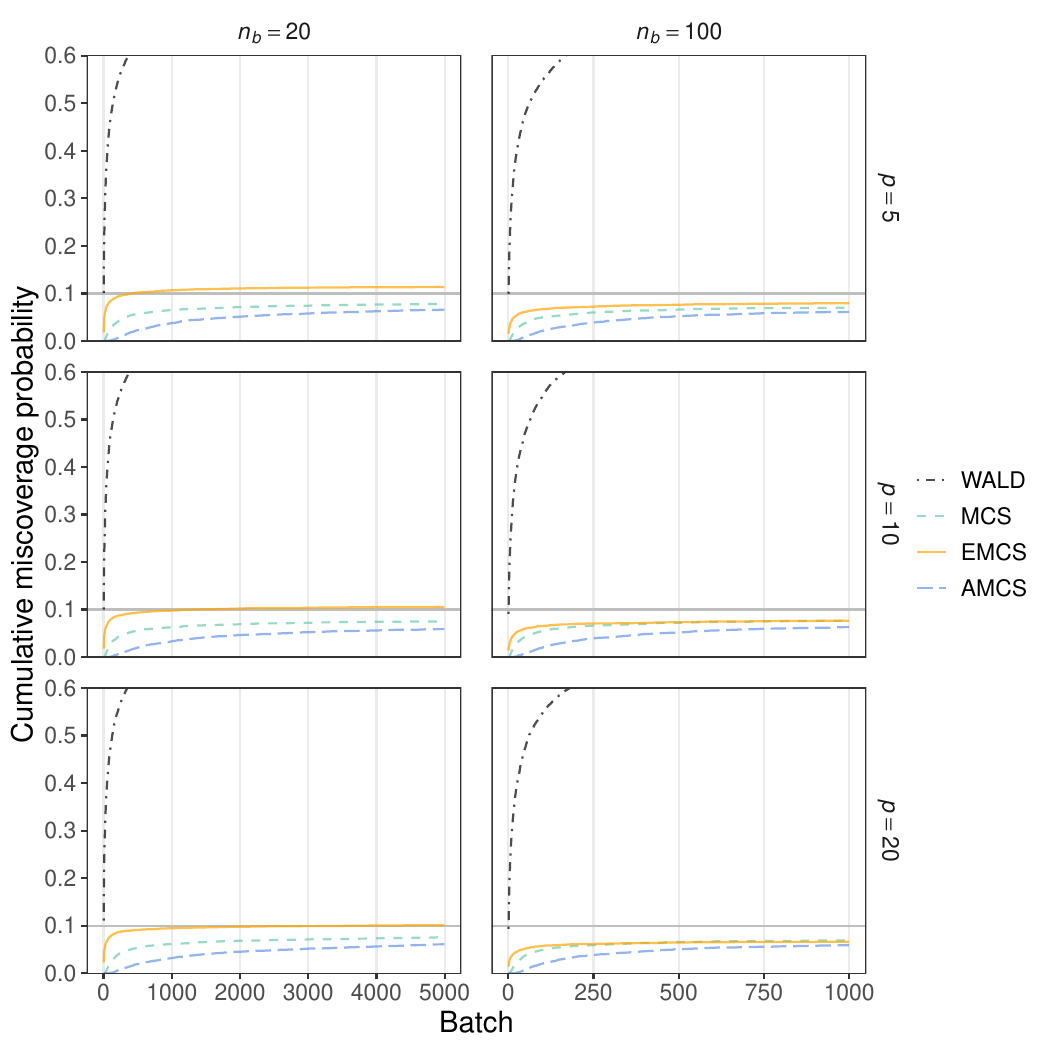}
		\caption{Inference about the coefficient $\beta_{k}=1.2$ of a continuous covariate in the logit regression model. Empirical cumulative null miscoverage probabilities of estimation regions with $\alpha=0.1$ for $p=5$ (top), $p=10$ (middle), $p=20$ (bottom), and $n_b=20$ (left), $n_b=100$ (right); WALD (dot-dashed), MCS (dashed), EMCS (solid) with $n_0=200$, AMCS (long-dashed) with $\rho$ optimized at $t=2 \times 10^4$. In each panel, the gray horizontal line marks the nominal value $\alpha$.  }\label{sim_Logit_Cont}	
	\end{center}
\end{figure}

The second experiment focuses on logistic regression models with 
$p \in \{ 5, 10, 20\}$ and $n_b \in \{20, 100\}$. 
In each setting, $n_0$ + $10^5$  observations, with $n_0=200$,  were simulated using the same procedure described for the numerical illustration in Figure \ref{sampleCSsBin}. 
The number of binary covariates for every $p$ was determined by rounding $(p-1)/2$ to the nearest even integer. 
Figures \ref{sim_Logit_Bin} and \ref{sim_Logit_Bin_NN} report estimated cumulative null and non-null miscoverage probabilities, respectively, of CSs for the coefficient of one binary covariate. Null miscoverage of EMCS is uniformly closest to the nominal $\alpha$, but AMCS exhibits better non-null miscoverage properties.
A similar pattern for the cumulative null miscoverage probabilities is shown in Figure \ref{sim_Logit_Cont}, which refers to the coefficient $\beta_{k}=1.2$ of one continuous covariate. 
When $n_b=20$ and $p \in \{5,10\}$ EMCS has empirical null cumulative miscoverage rate that slightly exceeds $\alpha$, but it gets more conservative as $n_b$ and $p$ increase.
Interestingly, this tendency is confirmed if we control for the number of miscoverage checks along the sequence (see Figures 1S--2S in Section 4 of the Supplementary Material).
Figures \ref{sim_Logit_Cont_NN} and \ref{sim_Logit_Cont_NN1} show empirical miscoverage probabilities for two different non-null 
 parameter values,  $\beta_k'=1.8$ and $\beta_k'=1$ respectively. The empirical tuning of the weight function in EMCS seems particularly beneficial in the latter, more challenging setting. AMCS performs best in both scenarios, although this finding relies crucially on the optimal choice of $\rho$.
 

\begin{figure}
	\begin{center}
		\includegraphics[height=16cm]{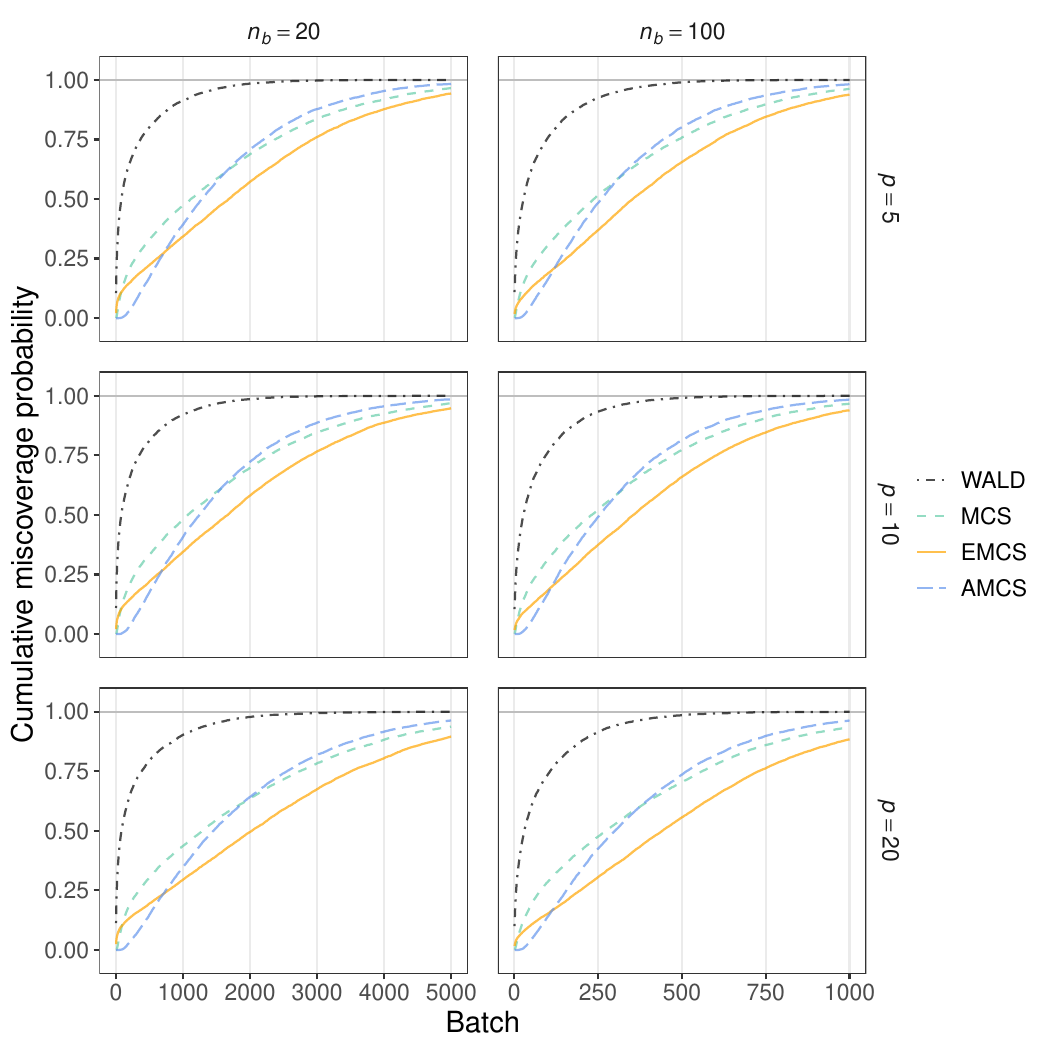}
		\caption{Inference about the coefficient $\beta_{k}=1.2$ of a continuous covariate in the logit regression model. Empirical cumulative non-null miscoverage probabilities, $\beta_k'=1.8$, of estimation regions with $\alpha=0.1$ for $p=5$ (top), $p=10$ (middle), $p=20$ (bottom), and $n_b=20$ (left), $n_b=100$ (right); WALD (dot-dashed), MCS (dashed), EMCS (solid) with $n_0=200$,  AMCS (long-dashed) with $\rho$ optimized at $t=2 \times 10^4$. In each panel, the gray horizontal line marks the 1 target value. }\label{sim_Logit_Cont_NN}	
	\end{center}
\end{figure}		

\begin{figure}
	\begin{center}
		\includegraphics[height=16cm]{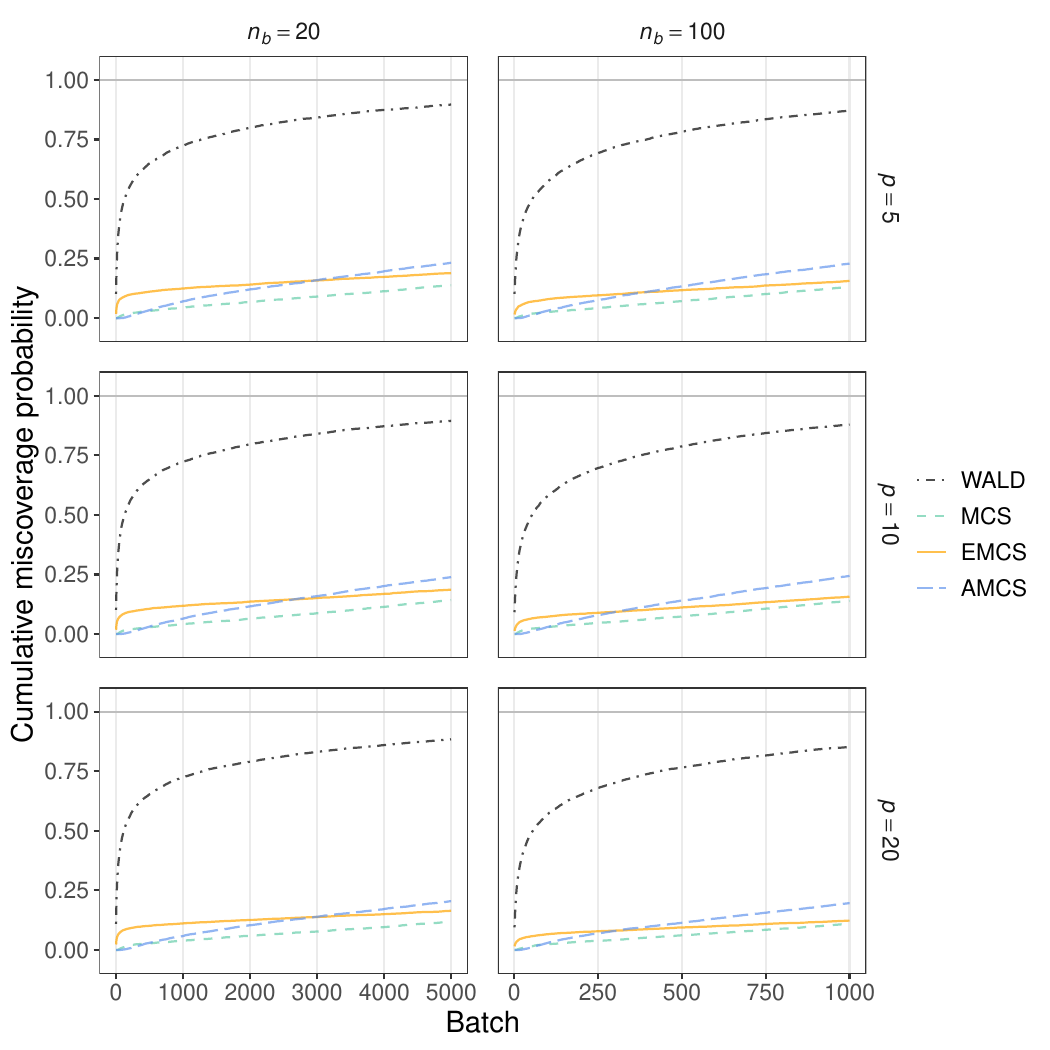}
		\caption{Inference about the coefficient $\beta_{k}=1.2$ of a continuous covariate in the logit regression model. Empirical cumulative non-null miscoverage probabilities, $\beta_k'=1$, of estimation regions with $\alpha=0.1$ for $p=5$ (top), $p=10$ (middle), $p=20$ (bottom), and $n_b=20$ (left), $n_b=100$ (right); WALD (dot-dashed), MCS (dashed), EMCS (solid) with $n_0=200$,  AMCS (long-dashed) with $\rho$ optimized at $t=2 \times 10^4$. In each panel, the gray horizontal line marks the 1 target value. }\label{sim_Logit_Cont_NN1}	
	\end{center}
\end{figure}		

\section{Data example}\label{sec:NASS}
We apply our strategy for efficiently computing mixture confidence sequences in GLMs to publicly available streaming data from the US National Automotive Sampling System (NASS) -- 
Crashworthiness data system\footnote{https://www.nhtsa.gov/national-automotive-sampling-system/crashworthiness-data-system}. In order to reproduce the analyses in \citet[Section 7]{luosong20}, we consider $B=84$ batches of monthly data spanning the seven-year period from January 2009 to December 2015, with a total sample size of $N_B=23\,185$ observations corresponding to drivers whose vehicle has been involved in an accident. The response in the assumed logistic model is a binary indicator of the event `Fatality' in the related crash, created from the ATREAT variable in the database. The included covariates are the driver's age, sex, seat belt use and drinking status, along with light condition, speed limit and presence of some traffic control function at the accident site. 
The variable Age is coded as a factor with three levels: young (less than 21 years), middle (21--64 years), old (at least 65 years). The variables Sex (=1 for males), Seat Belt, Drinking, Light Condition (=1 if different from daylight) and Traffic Control Function are binary, while Speed Limit records the statutory limit in km/h.

\begin{figure}
	\begin{center}
		\includegraphics[height=4.9in]{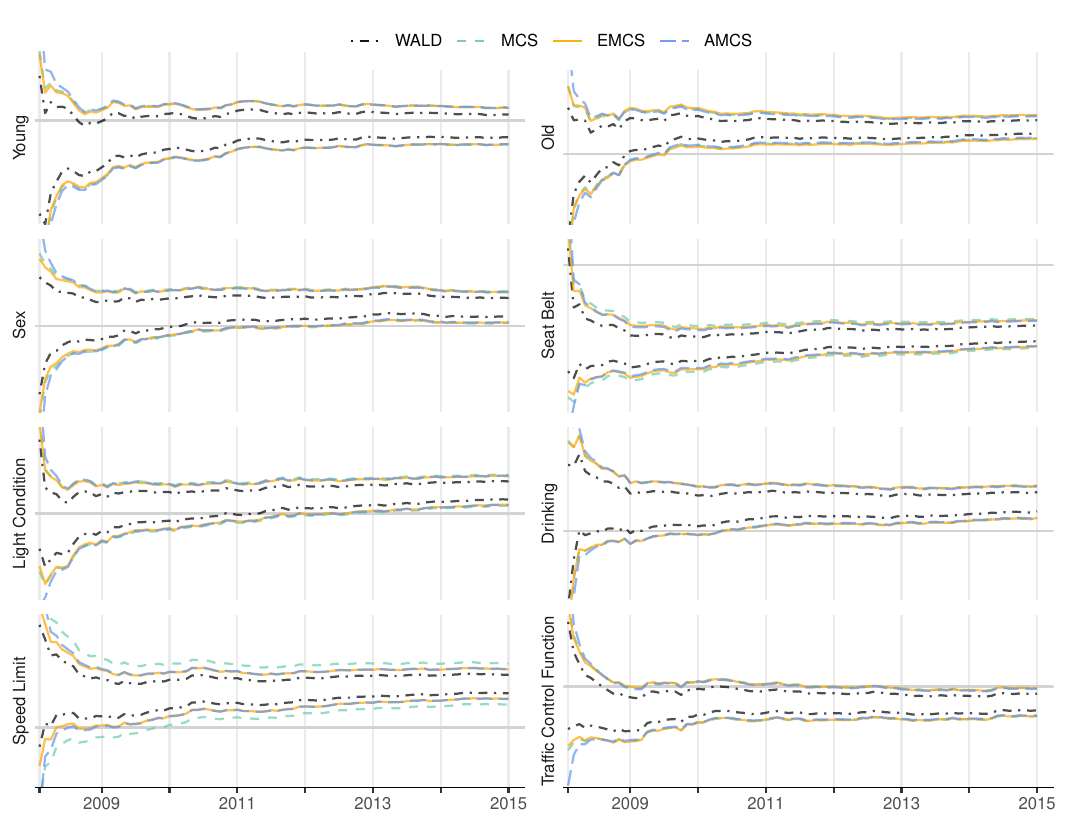}
		\caption{Inference about the regression coefficients in the logistic regression model for the NASS data via estimation regions with $\alpha=0.05$; WALD (dot-dashed), MCS (dashed), EMCS (solid) with $n_0=408$, AMCS (long-dashed) with $\rho$ optimized at $t=5000$. In each panel, the gray horizontal line marks the 0 reference value. }\label{figNASS}	
	\end{center}
\end{figure}

Figure \ref{figNASS} shows Wald confidence intervals and mixture CSs at the conventional 0.95 level for the eight coefficients of the logistic regression. The  parameters of the mixing distribution in EMCS are estimated using the first batch with $408$ observations, while the optimal $\rho$ in AMCS   is computed for $t = 5000$. We see that MCS is wider than the other two CSs for Speed Limit and Seat Belt, in line with the numerical example in Figure \ref{sampleCSsBin}, but usually the mixture CSs become practically indistinguishable from one another after the first year. They remain wider than the corresponding pointwise Wald-type intervals, which may affect conclusions about covariate effects on the response. Here, in particular, anytime-valid inferences provide little to no evidence of a significant association between Sex or Traffic Control Function and the probability of a fatal vehicle crash, contradicting standard fixed-sample-size
procedures.

\section{Discussion} \label{sec:final}
The performance of Wald-type confidence intervals in our simulations shows that it can be very dangerous to repeatedly apply fixed-sample-size inferential procedures in sequential data analyses. 
We have presented a simple and computationally efficient way to obtain mixture CSs for regression coefficients in GLMs under this scenario.
The considered  CSs exhibit an empirical null miscoverage probability which is close to the nominal level $\alpha$, and uniformly smaller for MCS and AMCS. When the tuning parameter $\rho$ is chosen wisely, the latter appears preferable for constructing estimation regions that are not too wide but stay reliable as new data batches accrue.
The application of the proposed strategy to the NASS data confirms the computational convenience of our approach, and the importance of drawing inferential conclusions based on anytime-valid tools that take the streaming nature of the observations into account.
In future works, a straightforward extension of our approach could build on the renewable estimation framework developed for high-dimensional GLMs in \cite{luo:23}.
%
%
%

\section{Supplementary material}
The repository https://github.com/cdicaterina/CSsGLMs provides the R scripts to reproduce all the analyses and outputs in the manuscript and includes some additional numerical results.

\section{Acknowledgments}
The authors thank the organizers and participants of the 3rd Workshop on Game-Theoretic Statistics and Sequential, Anytime-Valid Inference (SAVI), hosted by BIRS at the Chennai Mathematical Institute in India, for suggestions and discussions which helped improve this work.

%

\bibliographystyle{chicago}
\bibliography{CSglm_ref}

\end{document}